\global\def\draftcontrol{0}

   \def\versionno{ higher-derivative-conformal-order}

\catcode`\@=11

\expandafter\ifx\csname draftcontrol\endcsname\relax\global\def\draftcontrol{0}
\fi

{\count255=\time\divide\count255 by 60
\xdef\hourmin{\number\count255}
\multiply\count255 by-60\advance\count255 by\time
\xdef\hourmin{\hourmin:\ifnum\count255<10 0\fi\the\count255}}
\def\draftdate{\number\month/\number\day/\number\year\ \ \ \hourmin }

\newcommand\makepapertitle{\par
  \begingroup
    \renewcommand\thefootnote{\@fnsymbol\c@footnote}%
    \def\@makefnmark{\rlap{\@textsuperscript{\normalfont\@thefnmark}}}%
    \long\def\@makefntext##1{\parindent 1em\noindent
            \hb@xt@1.8em{%
                \hss\@textsuperscript{\normalfont\@thefnmark}}##1}%
     \newpage
     \global\@topnum\z@   
     \@makepapertitle
     \thispagestyle{empty}\@thanks
  \endgroup
  \setcounter{footnote}{0}%
  \global\let\thanks\relax
  \global\let\makepapertitle\relax
  \global\let\@makepapertitle\relax
  \global\let\@thanks\@empty
  \global\let\@author\@empty
  \global\let\@date\@empty
  \global\let\@title\@empty
  \global\let\title\relax
  \global\let\author\relax
  \global\let\date\relax
  \global\let\and\relax
  \def\version{\let\version\@version\@gobble}
}
\def\@makepapertitle{%
  \newpage
   \ifnum\draftcontrol=1 {}
   \version\versionno
   \vskip 3em%
   \else
   \hfill\hbox to 3cm {\parbox{4cm}{\@pubnum}\hss}%
   \vskip 3em%
   \fi
   \begin{center}%
   \let \footnote \thanks
     {\LARGE {\@title}}%
     \vskip 1.5em%
     {\normalsize
       \lineskip .5em%
       \begin{tabular}[t]{c}%
         \@author
       \end{tabular}\par}%
     \vskip 1.5em%
     {\@bstract}%
     \end{center}%
     \vskip 1.5em
     \@date%
   \par
}

\gdef\@pubnum{}
\def\pubnum#1{%
  \gdef\@pubnum{#1}}

\gdef\@bstract{}
\def\Abstract#1{%
  \gdef\@bstract{%
   \parbox{\textwidth-0pc}{%
   \centerline{\bf Abstract}\penalty1000%
\kern.2cm%
\noindent
\renewcommand\baselinestretch{1.0}%
{#1}}}
}

\def\ps@paper{\let\@mkboth\@gobbletwo%
     \ifnum\draftcontrol=1
    \def\@oddfoot{\hbox to \textwidth{\tiny \versionno \hfil\tiny\draftdate}%
    \hskip -\textwidth \hbox to \textwidth{\hfil\rm\thepage\hfil}}%
     \else\def\@oddfoot{\hbox to \textwidth{\hfil\rm\thepage\hfil}}
     \fi
     \let\@evenfoot\@oddfoot
}

\def\body{\clearpage
          \pagestyle{paper}
    }

\def\@version#1{\ifnum\draftcontrol=1
\typeout{}\typeout{#1}\typeout{}
\vskip3mm\centerline{\hbox{\fbox{\normalsize{\tt DRAFT -- #1 -- }
                   {\draftdate}}}}\vskip3mm
\fi}
\let\version\@version
\long\def\eqlabel#1{\ifnum\draftcontrol=1
                    \tag@false  
                    \tag*{(\theequation) \hbox to -0.2cm{\hspace{0cm}\small{#1}\hss}}
                    \refstepcounter{equation}
                    \edef\@currentlabel{\theequation}
                    \ltx@label{#1}          
                    \else
                    \label{#1}
                    \fi
                    }
\let\st@bibitem\@bibitem
\let\st@lbibitem\@lbibitem
\ifnum\draftcontrol=1
  \def\@bibitem#1{%
    \st@bibitem{#1}\a@@label{#1}\ignorespaces}
  \def\@lbibitem[#1]#2{%
    \st@lbibitem[#1]{#2}\a@@label{#2}\ignorespaces}
  \def\a@@label#1{%
    \gdef\a@lab{\smash{\normalfont\small#1}}
    \ifvmode
      \if@inlabel
        \global\setbox\@labels\hbox{%
          \llap{\a@lab\let\a@lab\relax
                \kern\@totalleftmargin\kern\marginparsep}%
          \box\@labels}%
      \fi
    \fi}
\fi

\documentclass[12pt,letterpaper]{article}

\usepackage{amsmath,amssymb,array,calc,epsfig,rotating,bm}
\usepackage[sort]{cite}
\usepackage{graphicx}
\usepackage{psfrag,verbatim}
\usepackage{xcolor}
\usepackage{hyperref}

\ifnum\draftcontrol=1
\fi

\renewcommand\baselinestretch{1.25}
\renewcommand\section{\@startsection {section}{1}{\z@}%
                                   {-3.5ex \@plus -1ex \@minus -.2ex}%
                                   {2.3ex \@plus.2ex}%
                                   {\normalfont\large\bfseries}}
\renewcommand\subsection{\@startsection{subsection}{2}{\z@}%
                                   {-3.25ex\@plus -1ex \@minus -.2ex}%
                                   {1.5ex \@plus .2ex}%
                                   {\normalfont\normalsize\bfseries}}
\renewcommand\subsubsection{\@startsection{subsubsection}{3}{\z@}%
                                   {-3.25ex\@plus -1ex \@minus -.2ex}%
                                   {1.5ex \@plus .2ex}%
                                   {\normalfont\normalsize\it}}
\renewcommand\paragraph{\@startsection{paragraph}{4}{\z@}%
                                   {-3.25ex\@plus -1ex \@minus -.2ex}%
                                   {1.5ex \@plus .2ex}%
                                   {\normalfont\normalsize\bf}}

\numberwithin{equation}{section}

\def\revise#1       {\raisebox{-0em}{\rule{3pt}{1em}}%
                     \marginpar{\raisebox{.5em}{\vrule width3pt\
                     \vrule width0pt height 0pt depth0.5em
                     \hbox to 0cm{\hspace{0cm}{%
                     \parbox[t]{4em}{\raggedright\footnotesize{#1}}}\hss}}}}

\newcommand\nxt[1]  {\\\fnxt#1}
\newcommand{\ie}{{\it i.e.,}\ }

\def\calb         {{\cal B}}
\def\calc         {{\cal C}}

\def\cale         {{\cal E}}
\def\calf         {{\cal F}}

\def\cali         {{\cal I}}

\def\call         {{\cal L}}
\def\calm         {{\cal M}}

\def\calo         {{\cal O}}

\def\reals        {{\mathbb R}}
\def\zet          {{\mathbb Z}}

\def\del          {\partial}

\def\Im           {{\rm Im\hskip0.1em}}

\def\sqr#1#2{{\vcenter{\vbox{\hrule height.#2pt
 \hbox{\vrule width.#2pt height#1pt \kern#1pt
 \vrule width.#2pt}\hrule height.#2pt}}}}

\def\dd{\delta}

\def\aa1{\phi}
\def\cc1{\psi}

\catcode`\@=12

\begin{document}


\title{\bf Holographic conformal order with higher derivatives}

\date{December 25, 2023}

\author{
Alex Buchel$^{1,2}$\\[0.4cm]
\it $^1$Department of Physics and Astronomy\\ 
\it University of Western Ontario\\
\it London, Ontario N6A 5B7, Canada\\
\it $^2$Perimeter Institute for Theoretical Physics\\
\it Waterloo, Ontario N2J 2W9, Canada\\
}

\Abstract{Conformal order are isotropic and translationary invariant thermal
states of a conformal theory with nonzero expectation value of certain
operators.  While ubiquitous in bottom-up models of holographic CFTs,
conformal order states are unstable in theories dual to bulk
two-derivative gravity. We explore conformal order in strongly coupled
theories with gravitational holographic duals involving higher
derivative corrections.
}

\makepapertitle

\body

\version\versionno
\tableofcontents

\section{Introduction and summary}\label{intro}

Gauge theory/gravity correspondence \cite{Maldacena:1997re,Aharony:1999ti}
has been a valuable tool in our understanding of strongly coupled matter.
It often lead to discoveries of new and unexpected phenomena. One such
discovery was a construction of {\it exotic hairy black holes} \cite{Buchel:2009ge},
predicting symmetry broken phases of strongly coupled gauge theories
that persist to arbitrary high temperatures. These holographic symmetry
broken phases exist even for AdS/CFT duals
\cite{Buchel:2020thm,Buchel:2020xdk,Buchel:2022zxl} and present
a holographic realization of a {\it conformal order}
\cite{Chai:2020zgq,Chaudhuri:2020xxb,Chai:2021djc,Chaudhuri:2021dsq}.
Specifically, for a CFT$ _{4}$ with a global
symmetry group $G$ in Minkowski space-time $\reals^{3,1}$
the existence of the
ordered phase implies that there are at least two distinct thermal
phases:
\begin{equation}
\frac{\calf}{T^{4}}=-\calc\ \times\
\begin{cases}
1,\qquad T^{-\Delta}\langle\calo_\Delta\rangle=0\ \Longrightarrow\
G\ {\rm is\ unbroken}; \\
\kappa,\qquad T^{-\Delta}\langle\calo_\Delta\rangle=\gamma\ne 0\ \Longrightarrow\
G\ {\rm is\ spontanuously\ broken},
\end{cases}
\eqlabel{phd}
\end{equation}
where $\calf$ is the free energy density, $T$ is the temperature,
$\calc$ is a positive constant
proportional to the central charge of the theory, and $\calo_\Delta$ is the
local order parameter for the symmetry breaking of conformal dimension
$\Delta$. The parameters $\kappa$ and $\gamma$
characterizing the thermodynamics of the
symmetry broken phase are  necessarily
constants\footnote{Direct CFT computation of
the values $\{\kappa,\gamma\}$ was never implemented.}. 
In all holographic constructions $\kappa$ was found to
be positive, implying that the symmetry broken phases are
thermodynamically stable. It was also found that in all  
models with two-derivative holographic duals $\kappa<1$,
implying that the symmetry broken phases are subdominant
both in the canonical and microcanonical ensembles.
The fact that the symmetry broken phase in the microcanonical ensemble is less entropic than
the symmetry preserving phase suggests that is must be
unstable \cite{Buchel:2005nt}. While there are no instabilities
in the hydrodynamic sector of the strongly coupled conformal order plasma, there is
an instability in the non-hydrodynamic sector --- one finds a quasinormal mode
of the dual hairy black brane with $\Im [\omega]>0$ \cite{Buchel:2020jfs}.
This instability is not lifted by compactifying the ordered phase on a positive curvature
spatial manifold, such as $S^3$ \cite{Buchel:2021ead}.

In this paper we continue pursuit of stable conformal order
and consider holographic models with higher-derivative gravitational duals.
Weyl$ ^4$ higher-derivative corrections to bulk Einstein gravity encode
finite 't Hooft coupling corrections of the dual conformal gauge theory
\cite{Gubser:1998nz}, and Riemann$ ^2$ corrections describe finite-$N$
effects in the dual theory \cite{Blau:1999vz,Nojiri:1999mh}.
Corrections of both types modify the relation between the
entropy densities of the boundary CFTs  and the horizon area densities of the
bulk black branes. This allows to engineer models where
only the symmetry broken phase triggers the higher-derivative corrections,
potentially increasing $\kappa$. We now review this idea, originally proposed in
\cite{Buchel:2010wf}.

Consider a five-dimensional theory of Einstein gravity in AdS, coupled to a scalar of
mass\footnote{From now on we set the radius of AdS to $L=1$.}
$m^2 L^2=\Delta(\Delta-4)$,
\begin{equation}
\begin{split}
S_5&=\frac{1}{16\pi G_N}\int_{\calm_5}d^5x \sqrt{-g}\ L_5
\equiv \frac{1}{16\pi G_N}\int_{\calm_5}d^5x \sqrt{-g}\ \biggl[
R+12-\frac{1}{2}(\del \phi)^2-\frac{m^2}{2}\phi^2\biggr]\,,
\end{split}
\eqlabel{2der0}
\end{equation}
where the bulk scalar $\phi$ is dual to an operator $\calo_\Delta$ of a
conformal dimension $\Delta$,
of a boundary CFT with a central charge $c=\frac{\pi}{8 G_N}$.
Note that the theory has a global $G\equiv\zet_2$ symmetry, $\phi\leftrightarrow -\phi$.
The only thermal states of the boundary CFT described by \eqref{2der0} are $AdS_5$
Schwarzschild black branes, \ie  the thermal expectation value of the operator
$\calo_\Delta$ vanishes,  leaving the global symmetry $G$ unbroken. 
The construction of the holographic conformal order proposed in
\cite{Buchel:2020xdk} relies on models where the scalar
has a nontrivial potential $V[\phi]$ instead, with the leading nonlinear
correction unbounded from below\footnote{This is actually
a very common feature in various (including top-down) holographic models.}.
As a simplest example we take $\Delta=3$ and 
\begin{equation}
V[\phi]=\frac{m^2}{2}\phi^2-b\phi^4=-\frac 32 \phi^2-b\phi^4\,, 
\eqlabel{defv}
\end{equation}
where $b>0$ is a bulk coupling, leading to
\begin{equation}
\begin{split}
S_5&= \frac{1}{16\pi G_N}\int_{\calm_5}d^5x \sqrt{-g}\ \biggl[
R+12-\frac{1}{2}(\del \phi)^2-V[\phi]\biggr]\,.
\end{split}
\eqlabel{2dern}
\end{equation}
The claim of \cite{Buchel:2020xdk} is that the thermal conformal order
in the model \eqref{2dern} always
exists in the limit $b\to +\infty$, where the thermal ordered phase is
holographically realized as an AdS-Schwarzschild black brane, with perturbatively
small ``scalar hair'',  
\begin{equation}
\phi=\frac{1}{\sqrt{b}}\biggl[p_0+\calo\left(\frac 1b\right)\biggr]
\qquad \Longleftrightarrow\qquad T^{-3}\langle\calo_3\rangle
=\gamma\propto \frac{1}{\sqrt{b}}\,,
\end{equation}
where $\gamma$ is the normalizable coefficient of $p_0$ near the $AdS_5$ boundary.

\begin{figure}[ht]
\begin{center}
\psfrag{x}[c]{{$r$}}
\psfrag{z}[cb]{{$p_{0}$}}
  \includegraphics[width=4.0in]{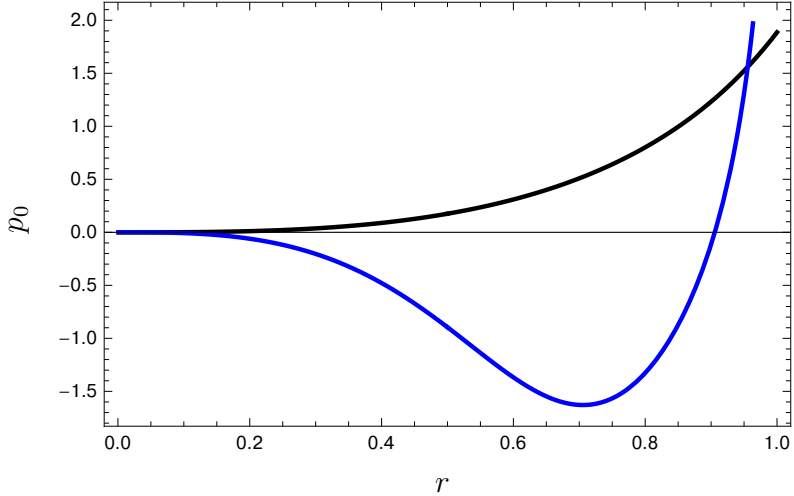}
\end{center}
 \caption{Condensation of the bulk scalar
 $\phi(r)=\frac{p_0(r)}{\sqrt{b}}$ in AdS-Schwarzschild background leads to
 perturbative in the limit $b\to +\infty$ conformal order in the holographic
 model \eqref{2dern}. The AdS black brane radial coordinate $r\in [0,1]$
runs from the boundary to the horizon. Different $p_0$ profiles,
the black and the blue curves, represent distinct conformal order
phases. 
}\label{figure1}
\end{figure}

The physical origin of the conformal ordered phase is easy to see. Notice that to leading
nonlinear order $\calo(b^{-1})$,
\begin{equation}
V[\phi]\equiv \frac{m_{eff}^2}{2}\ \phi^2\,,\qquad m_{eff}^2=
\Delta(\Delta-4)-2b\phi^2 =
\underbrace{\Delta(\Delta-4)-2p_0^2}_{\calo(b^0)}+\calo(b^{-1})\,.
\eqlabel{vlead}
\end{equation}
Thus, in the limit $b\to +\infty$, the effective mass of the bulk scalar
$\phi$ is shifted due to nonlinear negative quartic term in \eqref{defv}.
Potentially, when evaluated
at the AdS-Schwarzschild horizon\footnote{To leading order in $b$ the bulk
geometry is not modified.} it can dip below the Breitenlohner-Freedman bound,
triggering the instability and leading to 'hair'. This is precisely what we find:
in fig.~\ref{figure1} we present different 
profiles of  $p_0$ in the model \eqref{2dern}, realizing distinct conformal
order phases. The bulk radial coordinate $r\in[0,1]$,
with $r\to 0$ being the asymptotic $AdS_5$ boundary, and $r\to 1$
being the regular black brane horizon. Taking into account the bulk scalar
backreaction, we compute\footnote{See section \ref{tech} for details.}
the thermodynamic coefficient $\kappa$
in \eqref{phd} for different profiles reported in fig.~\ref{figure1},
\begin{equation}
\kappa=1+\frac 1b\cdot
\begin{cases}
-1.0(8)\,,\ &{\rm when}\ p_0= \mathbf{\color{black} p_0}\\
-21.3(8)\,,\ &{\rm when}\ p_0= \mathbf{\color{blue} p_0}
\end{cases}\qquad+\qquad\calo(b^{-2})\,.
\eqlabel{kappas0}
\end{equation}
Note that in both ordered phases $\kappa<1$, so that they are subdominant
relative to the symmetry preserving phase with $p_0\equiv 0$ and $\kappa=1$.

\begin{figure}[ht]
\begin{center}
\psfrag{x}[c]{{$\alpha$}}
\psfrag{y}[cb]{{$\kappa_{[2]}$}}
\psfrag{z}[ct]{{$|\alpha|\cdot\kappa_{[2]}$}}
  \includegraphics[width=3.0in]{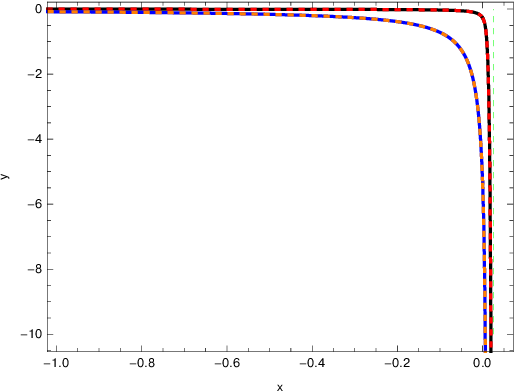}
  \includegraphics[width=3.0in]{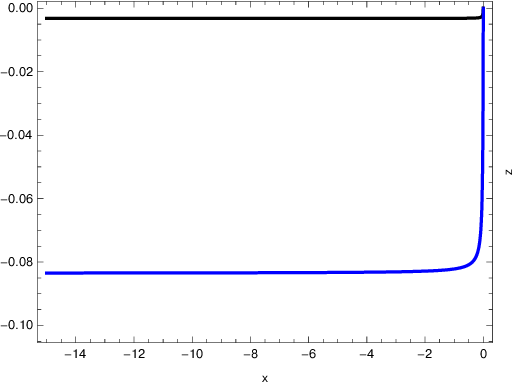}
\end{center}
 \caption{Leading corrections $\kappa_{[2]}$ to the conformal order thermodynamic parameter
 $\kappa=1+\frac 4b\cdot \kappa_{[2]}+\calo(b^{-2})$, as a function of the bulk coupling $\alpha$
 in the higher-derivative holographic model $\dd\call_2$. The solid black and blue
 curves correspond to scalar profiles without a root or with a single zero in the bulk,
 as in fig.~\ref{figure1}. The dashed red and orange curves represent the corresponding
 values of $\hat\kappa_{[2]}$. The vertical green dashed line (the left panel)
 indicates a critical value of $\alpha_{crit}=\frac{1}{40}$, beyond which the
 conformal order phases  do not exit.
}\label{figure2}
\end{figure}

\begin{figure}[ht]
\begin{center}
\psfrag{x}[c]{{$\alpha$}}
\psfrag{y}[cb]{{$\kappa_{[4]}$}}
\psfrag{z}[ct]{{$|\alpha|\cdot\kappa_{[4]}$}}
  \includegraphics[width=3.0in]{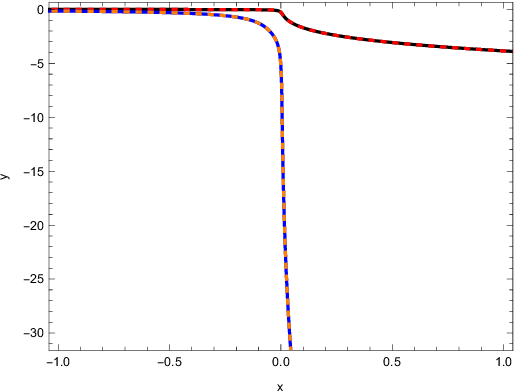}
  \includegraphics[width=3.0in]{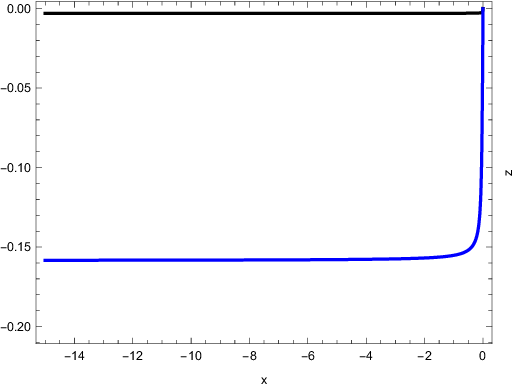}
\end{center}
 \caption{Leading corrections $\kappa_{[4]}$ to the conformal order thermodynamic parameter
 $\kappa=1+\frac 4b\cdot \kappa_{[4]}+\calo(b^{-2})$, as a function of the bulk coupling $\alpha$
 in the higher-derivative holographic model $\dd\call_4$. The solid black and blue
 curves correspond to scalar profiles without a root or with a single zero in the bulk,
 as in fig.~\ref{figure1}. The dashed red and orange curves represent the corresponding
 values of $\hat\kappa_{[4]}$. 
}\label{figure3}
\end{figure}

Motivated by \cite{Buchel:2010wf}, we modify the model \eqref{2dern}, including the
higher derivative corrections $\dd\call$,
\begin{equation}
\begin{split}
S_5&= \frac{1}{16\pi G_N}\int_{\calm_5}d^5x \sqrt{-g}\ \biggl[
R+12-\frac{1}{2}(\del \phi)^2-V[\phi]-b\phi^4\cdot\alpha\cdot \dd\call\biggr]\,,
\end{split}
\eqlabel{2der}
\end{equation}
for a constant parameter $\alpha$.
In particular, in this paper we consider two classes of
models\footnote{Throughout the paper we keep the subscripts $ _2$ or $ _4$ in
reference to models \eqref{dl2} and \eqref{dl4}.}
:
\begin{itemize}
\item
four-derivative curvature corrections described by: 
\begin{equation}
\dd\call_2\equiv R_{\mu\nu\rho\lambda} R^{\mu\nu\rho\lambda}\,;
\eqlabel{dl2}
\end{equation}
\item
eight-derivative curvature corrections described by: 
\begin{equation}
\dd\call_4\equiv  
C^{hmnk} C_{pmnq} C_h\ ^{rsp}C^q\ _{rsk}+\frac 12 C^{hkmn}C_{pqmn}
C_h\ ^{rsp}C^q\ _{rsk}\,,
\eqlabel{dl4}
\end{equation}
where $C$ is the Weyl tensor.
\end{itemize}
Note that the coupling constant of the higher-derivative term $\dd\call$
is small in the limit $b\to +\infty$ since
\begin{equation}
b \phi^4 = b\cdot \biggl(\frac{p_0}{\sqrt{b}}+\calo(b^{-3/2})\biggr)^4 \ \propto\ \frac{1}{b}\,.
\eqlabel{w4r2}
\end{equation}
Similar to \eqref{vlead},
\begin{equation}
V[\phi]+b\phi^4\cdot\alpha\cdot \dd\call\equiv \frac{m_{eff}^2}{2}\ \phi^2\,,\qquad m_{eff}^2=
\underbrace{\Delta(\Delta-4)-2 p_0^2\left(1-\alpha\cdot \dd\call\right)}_{\calo(b^0)}+\calo(b^{-1})\,.
\eqlabel{vlead2}
\end{equation}
Evaluated at the horizon of the AdS-Schwarzschild black brane both $\dd\call_2$ and $\dd\call_4$
are positive\footnote{See eqs. \eqref{di2} and \eqref{di4}.}.
Thus, we expect that $\alpha<0$ would facilitate the condensation
of the bulk scalar; additionally, precisely for $\alpha<0$ the Wald entropy \cite{Wald:1993nt} density 
of the higher-derivative black brane is larger that its Bekenstein entropy\footnote{See the relevant
discussion in section 2.1 of \cite{Buchel:2023fst}.},
further increasing the value of $\kappa$ in the ordered
phase.

We delegate the technical details of the analysis of model \eqref{2der}
to section \ref{tech}, and report the results only. We can extract
the thermodynamic parameter $\kappa$ in \eqref{phd} independently, either evaluating
the entropy density $s$, or the energy density $\cale$, of the corresponding
higher-derivative black brane solution, 
\begin{equation}
\frac{G_N}{\pi^3}\cdot \frac{s}{T^3}=\frac \kappa4\,,\qquad
\frac{G_N}{\pi^3}\cdot \frac{4\cale}{3T^4}=\frac {\hat\kappa}{4}\,.
\eqlabel{se24}
\end{equation}
The agreement $\kappa=\hat\kappa$ is a nontrivial consistency check on the computations.
\begin{itemize}
\item In the higher-derivative holographic model $\dd\call_2$ the conformal
order exists for $\alpha\in (-\infty,\frac{1}{40})$. We parameterize $\kappa=\kappa(\alpha)$
to order $\calo(b^{-1})$ as
\begin{equation}
\frac \kappa4=\frac14+\frac 1b\cdot \kappa_{[2]}(\alpha)+\calo(b^{-2})  \,,
\eqlabel{defk20}
\end{equation}
and consider the bulk scalar profiles as in fig.~\ref{figure1}. 
We find that, as predicted above --- see the solid black and blue curves in
fig.~\ref{figure2} --- 
\begin{equation}
\kappa_{[2]}(\alpha)\bigg|_{\alpha<0}\ >\ \kappa_{[2]}\bigg|_{\alpha=0}\,.
\eqlabel{k2in}
\end{equation}
However, $\kappa_{[2]}$ remains negative whenever the  conformal order phases exist.
In the limit $\alpha\to -\infty$, $\kappa_{[2]}\propto -\frac{1}{|\alpha|}$,
see the right panel of fig.~\ref{figure2}. For $\alpha>0$ the
conformal order phases become very subdominant compare to the symmetry
preserving phase, and cease to exist as $\alpha\to \frac{1}{40}$,
represented by the dashed vertical green line in the left panel of fig.~\ref{figure2}.
The dashed red and orange curves (the left panel) represent the corresponding
values of $\hat\kappa_{[2]}$. We find that $(\kappa_{[2]}/\hat\kappa_{[2]}-1)\propto 10^{-7}$
or better.
\item In the higher-derivative holographic model $\dd\call_4$ the conformal
order exists for any value of $\alpha$. We parameterize $\kappa=\kappa(\alpha)$
to order $\calo(b^{-1})$ as
\begin{equation}
\frac \kappa4=\frac14+\frac 1b\cdot \kappa_{[4]}(\alpha)+\calo(b^{-2})  \,,
\eqlabel{defk40}
\end{equation}
and consider the bulk scalar profiles as in fig.~\ref{figure1}. 
We find that --- see the solid black and blue curves in
fig.~\ref{figure3} --- 
\begin{equation}
\kappa_{[4]}(\alpha)\bigg|_{\alpha<0}\ >\ \kappa_{[4]}\bigg|_{\alpha=0}\,.
\eqlabel{k4in}
\end{equation}
Once again, $\kappa_{[4]}$ is always  negative.
In the limit $\alpha\to -\infty$, $\kappa_{[4]}\propto -\frac{1}{|\alpha|}$,
see the right panel of fig.~\ref{figure3}. 
The dashed red and orange curves (the left panel) represent the corresponding
values of $\hat\kappa_{[4]}$. We find that $(\kappa_{[4]}/\hat\kappa_{[4]}-1)\propto 10^{-6}$
or better.
\end{itemize} 

As we see, at least for the models considered,
the higher-derivative corrections can not make
the holographic conformal order phases to dominate the symmetric
phase. While we have not done the stability analysis as in
\cite{Buchel:2020jfs}, we do expect that the higher-derivative
black branes with a scalar hair constructed here have an unstable
quasinormal mode. It is possible that the holographic multiverse
simply does not allow for a stable thermal conformal order.
It would be interesting to rigorously prove this in
full generality.

\section{Technical details}\label{tech}

In this section we collect the technical details, necessary to reproduce
the results reported in section \ref{intro}.
We heavily rely on a recent construction of the higher-derivative
AdS-Schwarzschild black brane solutions in \cite{Buchel:2023fst}.

\subsection{Black brane geometry dual to thermal states of the
boundary theory}\label{background}

The background geometry dual to a thermal equilibrium state
of a boundary gauge theory takes form
\begin{equation}
ds_5^2=-c_1^2\ dt^2+c_2^2\ d\bm{x}^2+ c_3^2\ dr^2\,,
\eqlabel{5metric}
\end{equation}
where $c_i=c_i(r)$, and additionally $\phi=\phi(r)$.
The radial coordinate is $r\in [0,r_h]$, with $r_h=1$ being the
location of the regular black brane horizon,
\begin{equation}
\lim_{r\to r_h} c_1=0\,.
\eqlabel{defrh}
\end{equation}
Notice that at this stage we do not fix the residual diffeomorphism
associated with the reparametrization of the radial coordinate. 

One can efficiently compute the background equations of motion
from the effective one dimensional action,
\begin{equation}
S_1=\frac{1}{16\pi G_N}\int_0^{r_h} dr \biggl[\cali-b \phi^4\cdot \alpha \cdot \dd\cali\biggr]\,,
\eqlabel{s1}
\end{equation}
obtained from the
evaluation of \eqref{2der} on the ansatz \eqref{5metric}.
Here ($'\equiv \frac{d}{dr}$),
\begin{equation}
\cali=c_1 c_2^3 c_3 \biggl(12
-\frac{2 c_1''}{c_1 c_3^2}-\frac{6 c_2''}{c_2 c_3^2}-\frac{6 (c_2')^2}{c_2^2 c_3^2}
+\frac{6 c_3' c_2'}{c_2 c_3^3}-\frac{6 c_1' c_2'}{c_2 c_1 c_3^2}
+\frac{2 c_1' c_3'}{c_1 c_3^3}-\frac{(\phi')^2}{2c_3^2}-V
\biggr)\,,
\eqlabel{i2der}
\end{equation}
with the higher derivative contributions in model  \eqref{dl2} given by
\begin{equation}
\begin{split}
&\dd\cali_2=\frac{c_1 c_2^3}{ c_3^3} \biggl(
4 \left[\frac{ c_1''}{c_1}-\frac{c_1'c_3'}{c_1c_3}\right]^2+12
\left[\frac{ c_2''}{c_2}-\frac{c_2'c_3'}{c_2c_3}\right]^2+
\frac{12 (c_1')^2 (c_2')^2}{c_2^2 c_1^2}
+\frac{12 (c_2')^4}{c_2^4}
\biggr)\,,
\end{split}
\eqlabel{di2}
\end{equation}
and in model \eqref{dl4}  by
\begin{equation}
\begin{split}
\dd\cali_4=\frac{5}{36} c_1 c_2^3 c_3 \biggl(
\frac{c_1' c_2'}{c_2 c_1 c_3^2}+\frac{c_1' c_3'}{c_1 c_3^3}
-\frac{(c_2')^2}{c_2^2 c_3^2}-\frac{c_3' c_2'}{c_2 c_3^3}
-\frac{c_1''}{c_1 c_3^2}+\frac{c_2''}{c_2 c_3^2}\biggr)^4\,.
\end{split}
\eqlabel{di4}
\end{equation}
From \eqref{s1} we obtain the following equations of
motion:
\begin{equation}
\begin{split}
&0=c_1''-\frac{c_1 (c_2')^2}{c_2^2}+\frac{2 c_2' c_1'}{c_2}
-\frac{c_3' c_1'}{c_3}+\frac{(\phi')^2}{12} c_1+\frac16 c_3^2 c_1 (V-12)+b\phi^4\cdot \alpha\cdot
\cale_1\,,
\end{split}
\eqlabel{eq1}
\end{equation}
\begin{equation}
\begin{split}
&0=c_2''-\frac{c_2' c_3'}{c_3}
+\frac{(c_2')^2}{c_2}+\frac{(\phi')^2}{12} c_2 +\frac16 c_2 c_3^2 (V-12)+b\phi^4\cdot \alpha\cdot
\cale_2\,,
\end{split}
\eqlabel{eq2}
\end{equation}
\begin{equation}
\begin{split}
&0=(\phi')^2 -\frac{12 (c_2')^2}{c_2^2}-\frac{12 c_2' c_1'}{c_2 c_1}
-2 c_3^2 (V-12)+b\phi^4\cdot \alpha\cdot
\cale_3\,,
\end{split}
\eqlabel{eq3}
\end{equation}
\begin{equation}
\begin{split}
&0=\phi''-\frac{\phi' c_3'}{c_3}+\frac{3 \phi' c_2'}{c_2}+\frac{c_1' \phi'}{c_1}-c_3^2\ \del V
+b\phi^3\cdot \alpha\cdot
\cale_4\,,
\end{split}
\eqlabel{eq4}
\end{equation}
where $\cale_j$ are functionals\footnote{For readability we will not present their explicit
expressions here.} of $\{c_i,\phi\}$ such that
\begin{equation}
\cale_j\left[c_i\,;\,\lambda\cdot \phi\right]=\cale_j\left[c_i\,;\,\phi\right]\,,
\eqlabel{edef}
\end{equation}
for a constant $\lambda$.
We verified that the constraint \eqref{eq3} is consistent with the remaining equations.

On-shell, \ie evaluated when \eqref{eq1}-\eqref{eq4} hold, the effective action \eqref{s1} is a total derivative. Specifically,
we find 
\begin{equation}
\cali-b\phi^4\cdot\alpha\cdot\dd\cali=-\frac{6c_2^2c_1}{c_3}\ \cdot\ {\rm eq.}\eqref{eq2}
+\frac{d}{dr}\biggl\{-\frac{2c_2^3 c_1'}{c_3} +b\phi^4\cdot\alpha\cdot \dd\calb
\biggr\}\,,
\eqlabel{totder}
\end{equation}
with the higher derivative terms $\dd\calb$ given by
\begin{equation}
\begin{split}
&\dd\calb_2=-\frac{8 c_2^3 c_1}{c_3^3} \biggl(
\frac{c_3'' c_1'}{c_3 c_1}
-\frac{c_1'''}{c_1}+\frac{2 c_1'' c_1'}{c_1^2}
-\frac{3 c_1''c_2'}{c_2 c_1}
+\frac{3 (c_2')^2 c_1'}{c_2^2 c_1}
+\frac{3 c_1'' c_3'}{c_3 c_1}
-\frac{2 c_3' (c_1')^2}{c_3 c_1^2}
+\frac{3 c_2' c_3' c_1'}{c_3 c_2 c_1}\\
&-\frac{3 (c_3')^2 c_1'}{c_3^2 c_1}
-\frac{4 c_1'' \phi'}{c_1 \phi}
+\frac{4 \phi' c_3' c_1'}{c_3 c_1 \phi}
\biggr)\,;
\end{split}
\eqlabel{defb2}
\end{equation}
and
\begin{equation}
\begin{split}
&\dd\calb_4=\frac{5c_1c_2^3}{9c_3^7}\biggl(
\frac{(c_2')^2}{c_2^2}+\frac{c_2' c_3'}{c_3 c_2}
-\frac{c_2''}{c_2}-\frac{c_2' c_1'}{c_2 c_1}
-\frac{c_3' c_1'}{c_3 c_1}+\frac{c_1''}{c_1}
\biggr)^2\ \cdot\
\biggl(
\frac{3 c_1'''}{c_1}
-\frac{3c_2'''}{c_2}-\frac{c_1''}{c_1}
\biggl[\frac{4c_1'}{c_1}-\frac{c_2'}{c_2}\\&+\frac{9 c_3'}{c_3}
-\frac{4 \phi'}{\phi}\biggr]
-\frac{c_2''}{c_2} \biggl[
\frac{2 c_1'}{c_1}-\frac{5 c_2'}{c_2}-\frac{9 c_3'}{c_3}
+\frac{4 \phi'}{\phi}\biggr]
+\left( \frac{c_1'}{c_1}-\frac{c_2'}{c_2}\right)
\biggl[
\frac{4 c_1' c_2'}{c_2 c_1}
+\frac{2 (c_2')^2}{c_2^2}+\frac{4 c_1' c_3'}{c_3 c_1}
\\&-\frac{3 c_3''}{c_3}+\frac{5 c_3' c_2'}{c_3 c_2}
+\frac{9 (c_3')^2}{c_3^2}
-\frac{4 c_2' \phi'}{c_2 \phi}
-\frac{4 c_3' \phi'}{c_3 \phi}
\biggr]
\biggr)\,.
\end{split}
\eqlabel{defb4}
\end{equation}

In what follows we will need the entropy density $s$,
the energy density $\cale$, and the temperature $T$ of the boundary thermal state.
The temperature is determined by requiring the vanishing of the conical deficit angle of the 
analytical continuation of the geometry \eqref{5metric},
\begin{equation}
2\pi T = \lim_{r\to r_h}\left[ -\frac{c_2}{c_3}\ \left(\frac{c_1}{c_2}\right)'\right]= \lim_{r\to r_h}
\left[-\frac{c_1'}{c_3}+\frac{c_1c_2'}{c_2 c_3}\right]=\lim_{r\to r_h}
\left[-\frac{c_1'}{c_3}\right]\,,
\eqlabel{deft}
\end{equation}
where to obtain the last equality we used \eqref{defrh}.
The thermal entropy density of the boundary gauge theory is identified with the
entropy density of the dual black brane \cite{Witten:1998zw}. Since our holographic model
contains higher-derivative terms, the Bekenstein entropy $s_B$,
\begin{equation}
s_B=\lim_{r\to r_h} \frac{c_2^3}{4 G_N}\,,
\eqlabel{sb}
\end{equation}
must be replaced with the Wald entropy $s_W$ \cite{Wald:1993nt},
\begin{equation}
s_W=-\frac{1}{8\pi G_N}\lim_{r\to r_h}\biggl[ c_2^3\ \epsilon_{\mu\nu}\epsilon_{\rho\lambda}
\frac{\dd L_5}{\dd R_{\mu\nu\rho\lambda}} \biggr]\,,
\eqlabel{sw}
\end{equation}
\ie $s=s_W$.
The simplest way to compute the Wald entropy density is instead to use the boundary thermodynamics:
\nxt According to the holographic correspondence \cite{Maldacena:1997re,Aharony:1999ti}, the on-shell
gravitational action $S_1$, properly renormalized \cite{Skenderis:2002wp},
has to be identified with the boundary gauge theory free energy density
$\calf$ as follows,
\begin{equation}
-\calf=S_1\bigg|_{\rm on-shell} =\frac{1}{16\pi G_N} \int_0^{r_h}
dr\ \frac{d}{dr}\biggl\{-\frac{2c_2^3 c_1'}{c_3} +b\phi^4\cdot\alpha\cdot \dd\calb
\biggr\}  +\lim_{r\to 0}\ \biggl[S_{GH}+S_{ct}\biggr]\,,
\eqlabel{deff}
\end{equation}
where we used \eqref{totder}. $S_{GH}$ is a generalized Gibbons-Hawking term \cite{Buchel:2004di},
necessary to have a well-defined variational principle, and $S_{ct}$ is the counter-term
action. 
\nxt Eq.\eqref{deff} can be rearranged to explicitly implement the basic
thermodynamic relation $-\calf=s T-\cale$ between the free energy density $\calf$,
the energy density $\cale$ and the entropy density $s$ \cite{Buchel:2004hw}:
\begin{equation}
\begin{split}
-\calf=\frac{1}{16\pi G_N}\lim_{r\to r_h}\biggl[&-\frac{2c_2^3 c_1'}{c_3} +b\phi^4\cdot\alpha\cdot \dd\calb
\biggr]\\
&- \lim_{r\to 0}\biggl[\frac{1}{16\pi G_N}\left(
-\frac{2c_2^3 c_1'}{c_3} +b\phi^4\cdot\alpha\cdot \dd\calb
\right)+S_{GH}+S_{ct}\biggr]\,.               
\end{split}
\eqlabel{btr}
\end{equation}
\nxt From \eqref{btr} we identify\footnote{Strictly speaking, \eqref{defst} is correct
up to an arbitrary constant. But this constant must be set to zero from the
comparison with thermal AdS, in which case the black brane geometry
is dual to a thermal state of a boundary CFT with vanishing entropy
in the limit $T\to 0$.}
\begin{equation}
s T \equiv  s_W T = \frac{1}{16\pi G_N}\lim_{r\to r_h}\biggl[-\frac{2c_2^3 c_1'}{c_3}
+b\phi^4\cdot\alpha\cdot \dd\calb
\biggr]\,,
\eqlabel{defst}
\end{equation}
\begin{equation}
\begin{split}
\cale=&\lim_{r\to 0}\biggl[\frac{1}{16\pi G_N}\left(
-\frac{2c_2^3 c_1'}{c_3} +b\phi^4\cdot\alpha\cdot \dd\calb
\right)+\left\{S_{GH}\right\}+\left[S_{ct}\right]\biggr]\\
=&\frac{1}{16\pi G_N}\lim_{r\to 0}\biggl[\left(
-\frac{2c_2^3 c_1'}{c_3} +b\phi^4\cdot\alpha\cdot \dd\calb
\right)+\biggl\{-\frac{2(c_1c_2^3)'}{c_3}+\cdots\biggr\}+\left[ 6c_1c_2^3\right]\biggr]\,,
\end{split}
\eqlabel{defe}
\end{equation}
where in the second line $\{\cdots\}$ denote higher-derivative (generalized) GH terms that
vanish in the limit $r\to 0$; we also included the only relevant (non-vanishing
in the limit) counterterm in $[\ ]$. 

\subsection{Near-conformal conformal order}\label{ncco}

The near-conformal limit is achieved as $b\to +\infty$ \cite{Buchel:2020xdk},
\begin{equation}
\begin{split}
&c_1=\frac{\sqrt{1-r^4}}{r}\ \biggl(1+\sum_{i=1}^\infty \frac{ g_{1,i}(r)}{b^i}\biggr)^{1/2}\,,
\ \ c_2=\frac1r\,,\ \  c_{3}=\frac{1}{r\sqrt{1-r^4}}
\biggl(1+\sum_{i=1}^\infty \frac{ g_{2,i}(r)}{b^i}\biggr)^{-1}\,,\\
&\phi=\frac{1}{b^{1/2}}\ \sum_{i=0}^\infty \frac{p_i(r)}{b^i}\,.
\end{split}
\eqlabel{pert}
\end{equation}

\subsubsection{Model $\dd \call_2$ to leading order in $\frac 1b$}

Using \eqref{pert}, we find from \eqref{eq1}-\eqref{eq4}: 
\begin{equation}
\begin{split}
&0=p_0''+\frac{r^4+3 }{r (r^4-1)}\ p_0'+\frac{p_0}{(r^4-1) r^2} \biggl(
4 p_0^2\ (72 \alpha r^8-1+40 \alpha)+m^2\biggr)\,,
\end{split}
\eqlabel{eom21}
\end{equation}
where the last term $(\cdots+m^2)$ represents the {\it effective} scalar mass $m_{eff}^2$.
When $\alpha\ge \frac{1}{40}$, the effective mass is larger than $m^2$ for all values of $r$,
and the conformal order ceases to exist\footnote{We confirm this heuristics numerically.}.
\begin{equation}
\begin{split}
&0=g_{2,1}'+\frac{4 g_{2,1}}{r (r^4-1)}+\frac{r}{12} \biggl( 192 p_0^2\ \alpha (3 r^4-1)-1\biggr)\
(p_0')^2+\frac{16p_0^3 \alpha (9 r^8-14 r^4+1)}{3(r^4-1)}\ p_0'
\\&-\frac{p_0^2}{12r (r^4-1)} \biggl(
256 \alpha (3 r^4-1) (72 \alpha r^8-1+40 \alpha) p_0^4
+(48 \alpha r^8+192 \alpha m^2 r^4\\&-64 \alpha m^2+2-16 \alpha) p_0^2
-m^2\biggr)\,,
\end{split}
\eqlabel{eom22}
\end{equation}
\begin{equation}
\begin{split}
0=&g_{1,1}'+\frac{8 g_{2,1}}{r (r^4-1)}+\frac r6\ (p_0')^2 
+\frac{64p_0^3 \alpha (3 r^8+r^4-2)}{3(r^4-1)}\ (p_0')+\frac{p_0^2}{6r (r^4-1)} \biggl(
2 p_0^2 (-24 \alpha r^8\\&-1+8 \alpha)+m^2\biggr)\,,
\end{split}
\eqlabel{eom215}
\end{equation}
\begin{equation}
\begin{split}
&0=g_{1,1}''+\frac{3 (r^4+1)}{r (r^4-1)}\ g_{1,1}'-\frac{16 r^2 g_{2,1}}{(r^4-1)^2}
+\frac{r^4}{3(r^4-1)} \biggl(192 p_0^2\ \alpha (3 r^4-5)+1\biggr)\ (p_0')^2
\\&+\frac{128p_0^3 r^3 \alpha (6 r^8-17 r^4+13)}{3(r^4-1)^2}\  p_0'
-\frac{r^2 p_0^2}{3(r^4-1)^2} \biggl(
256 \alpha (3 r^4-5) (72 \alpha r^8-1\\&+40 \alpha) p_0^4
+(144 \alpha r^8+192 \alpha m^2 r^4-192 \alpha r^4-320 \alpha m^2
-2+16 \alpha) p_0^2+m^2\biggr) \,.
\end{split}
\eqlabel{eom23}
\end{equation}
It is easy to check that \eqref{eom23} is consistent with \eqref{eom21}-\eqref{eom215}.

The background equations \eqref{eom21}-\eqref{eom215} are solved subject to the following asymptotics:
\nxt near the AdS boundary, \ie as $r\to 0$, it is characterized by $\{p_{0;3},g_{2,1;4}\}$,
\begin{equation}
p_0=p_{0;3} r^3+\calo(r^7)\,,\qquad g_{2,1}=g_{2,1;4} r^4+\calo(r^6)\,,\qquad g_{1,1}=2g_{2,1;4} r^4+\calo(r^8)\,;
\eqlabel{2uv}
\end{equation}
\nxt in the vicinity of the black brane horizon, \ie as $y\equiv (1-r)\to 0$, it is characterized
by $\{p_{0;0}^h,g_{1,1;0}^h\}$,
\begin{equation}
\begin{split}
&p_0=p_{0;0}^h+\calo(y)\,,\qquad g_{1,1}=g_{1,1;0}^h+\calo(y)\,,\\
&g_{2,1}=\frac{(p_{0;0}^h)^2}{16}-\biggl(\frac{10}{3}\alpha
-\frac{1}{24}\biggr) (p_{0;0}^h)^4+\biggl(-\frac{16}{3}\alpha+\frac{1792}{3}\alpha^2\biggr) (p_{0;0}^h)^6
+\calo(y)\,.
\end{split}
\eqlabel{2ir}
\end{equation}

From \eqref{defst} and \eqref{defe} we extract:
\begin{equation}
\begin{split}
&\frac{G_N}{\pi^3}\cdot\frac{s}{T^3}=\frac\kappa4\equiv \frac 14+\frac 1b\cdot \kappa_{[2]}+
\calo(b^{-2})\,,\\
&\kappa_{[2]}=\biggl(4 \alpha-448 \alpha^2\biggr) (p_{0;0}^h)^6
+\biggl(\frac12 \alpha-\frac{1}{32} \biggr) (p_{0;0}^h)^4
-\frac{3}{64} (p_{0;0}^h)^2-\frac38 g_{1,1;0}^h\,;
\end{split}
\eqlabel{st2}
\end{equation}

\begin{equation}
\begin{split}
&\frac{G_N}{\pi^3}\cdot\frac{4\cale}{3T^4}=\frac{\hat\kappa}{4}\equiv \frac 14+\frac 1b\cdot
{\hat\kappa}^{[2]}+
\calo(b^{-2})\,,\\
&\hat{\kappa}^{[2]}=\biggl(\frac{16}{3} \alpha
-\frac{1792}{3} \alpha^2\biggr)
(p_{0;0}^h)^6+\biggl(\frac{10}{3} \alpha-\frac{1}{24}\biggr)
(p_{0;0}^h)^4
-\frac{1}{16} (p_{0;0}^h)^2\\&
-\frac12 g_{1,1;0}^h-\frac12 g_{2,1;4}\,.
\end{split}
\eqlabel{e2}
\end{equation}

\subsubsection{Model $\dd \call_4$ to leading order in $\frac 1b$}

Using \eqref{pert}, we find from \eqref{eq1}-\eqref{eq4}: 
\begin{equation}
\begin{split}
&0=p_0''+\frac{r^4+3}{r (r^4-1)}\ p_0'+\frac{p_0}{r^2 (r^4-1)} \biggl( 4 p_0^2\ (180 \alpha r^{16}-1)
+m^2\biggr)\,,
\end{split}
\eqlabel{eom41}
\end{equation}
\begin{equation}
\begin{split}
&0=g_{2,1}'+\frac{4 g_{2,1}}{r (r^4-1)}+\frac{r}{12} \biggl(2880 \alpha p_0^2 r^{12}-1\biggr)\
(p_0')^2+\frac{80 \alpha p_0^3 r^{12} (19 r^4-21)}{r^4-1}\ p_0'
\\&-\frac{p_0^2}{12r (r^4-1)} \biggl(
691200 p_0^4 \alpha^2 r^{28}+120 \alpha p_0^2 r^{12} (-171 r^4-32 p_0^2 +8 m^2+144)
+2 p_0^2 \\&-m^2\biggr)\,,
\end{split}
\eqlabel{eom42}
\end{equation}
\begin{equation}
\begin{split}
&0=g_{1,1}'+\frac{8 g_{2,1}}{r (r^4-1)}+\frac r6\ (p_0')^2 +\frac{320 p_0^3 \alpha r^{16}}{r^4-1}\
p_0'+\frac{p_0^2}{6r (r^4-1)} \biggl(2 p_0^2 (1620 \alpha r^{16}-1)\\&+m^2\biggr)\,,
\end{split}
\eqlabel{eom415}
\end{equation}
\begin{equation}
\begin{split}
&0=g_{1,1}''+\frac{3 (r^4+1)}{r (r^4-1)}\ g_{1,1}'-\frac{16 r^2 g_{2,1}}{(r^4-1)^2}
+\frac{r^4}{3(r^4-1)} \biggl(2880 p_0^2 \alpha r^8 (r^4-2)+1\biggr)\ (p_0')^2
\\&+\frac{640 \alpha p_0^3 r^{11} (10 r^8-30 r^4+21)}{(r^4-1)^2}\ p_0'-\frac{r^2 p_0^2}{3(r^4-1)^2}
\biggl(
691200 p_0^4 r^{24} (r^4-2) \alpha^2+120 p_0^2 r^8 (\\
&-189 r^8-32 p_0^2 r^4+8 m^2 r^4+504 r^4
+64 p_0^2 -16 m^2-288) \alpha-2 p_0^2 +m^2
\biggr) \,.
\end{split}
\eqlabel{eom43}
\end{equation}
It is easy to check that \eqref{eom43} is consistent with \eqref{eom41}-\eqref{eom415}.

The background equations \eqref{eom41}-\eqref{eom415} are solved subject to the following asymptotics:
\nxt near the AdS boundary, \ie as $r\to 0$, it is characterized by $\{p_{0;3},g_{2,1;4}\}$,
\begin{equation}
p_0=p_{0;3} r^3+\calo(r^7)\,,\qquad g_{2,1}=g_{2,1;4} r^4+\calo(r^6)\,,\qquad g_{1,1}=2g_{2,1;4} r^4+\calo(r^8)\,;
\eqlabel{4uv}
\end{equation}
\nxt in the vicinity of the black brane horizon, \ie as $y\equiv (1-r)\to 0$, it is characterized
by $\{p_{0;0}^h,g_{1,1;0}^h\}$,
\begin{equation}
\begin{split}
&p_0=p_{0;0}^h+\calo(y)\,,\qquad g_{1,1}=g_{1,1;0}^h+\calo(y)\,,\\
&g_{2,1}=\frac{(p_{0;0}^h)^2}{16}-\biggl(\frac{195}{2}\alpha
-\frac{1}{24}\biggr) (p_{0;0}^h)^4+\biggl(-40\alpha+7200\alpha^2\biggr) (p_{0;0}^h)^6
+\calo(y)\,.
\end{split}
\eqlabel{4ir}
\end{equation}

From \eqref{defst} and \eqref{defe} we extract:
\begin{equation}
\begin{split}
&\frac{G_N}{\pi^3}\cdot\frac{s}{T^3}=\frac\kappa4\equiv \frac 14+\frac 1b\cdot \kappa_{[4]}+
\calo(b^{-2})\,,\\
&\kappa_{[4]}=-30 \alpha (180 \alpha-1) (p_{0;0}^h)^6
+\biggl(-\frac{1}{32}+\frac{465}{8} \alpha\biggr) (p_{0;0}^h)^4-\frac{3}{64}
(p_{0;0}^h)^2-\frac38 g_{1,1;0}^h\,;
\end{split}
\eqlabel{st4}
\end{equation}

\begin{equation}
\begin{split}
&\frac{G_N}{\pi^3}\cdot\frac{4\cale}{3T^4}=\frac{\hat\kappa}{4}\equiv \frac 14+\frac 1b\cdot
{\hat\kappa}^{[4]}+
\calo(b^{-2})\,,\\
&\hat{\kappa}^{[4]}=(40  \alpha-7200 \alpha^2) (p_{0;0}^h)^6
+\biggl(\frac{195}{2} \alpha-\frac{1}{24} \biggr) (p_{0;0}^h)^4
-\frac{1}{16} (p_{0;0}^h)^2-\frac12 g_{1,1;0}^h\\&-\frac12 g_{2,1;4}\,.
\end{split}
\eqlabel{e4}
\end{equation}

\section*{Acknowledgments}
I would like to thank S.Cremonini and L.Early for collaboration on
\cite{Buchel:2023fst}.
Research at Perimeter
Institute is supported by the Government of Canada through Industry
Canada and by the Province of Ontario through the Ministry of
Research \& Innovation. This work was further supported by
NSERC through the Discovery Grants program.


\bibliographystyle{JHEP}
\bibliography{hdco}

\end{document}